\documentclass[aps,a4paper,superscriptaddress,showpacs,preprintnumbers,
amsmath,amssymb]{revtex4}
\usepackage{graphicx}
\usepackage{dcolumn}
\usepackage{color}
\usepackage{latexsym,amsfonts}
\usepackage{textcomp}
\usepackage{bm}

\usepackage{ulem}

\begin{document}

\title{GSI Oscillations\\ as Interference of Neutrino Flavour
  Mass--Eigenstates\\ and Measuring Process}

\author{A. N. Ivanov}\email{ivanov@kph.tuwien.ac.at}\affiliation{Atominstitut,
  Technische Universit\"at Wien, Stadionalle 2, A-1020 Wien,
  Austria}\author{P. Kienle}\affiliation{Stefan Meyer Institut f\"ur
  subatomare Physik \"Osterreichische Akademie der Wissenschaften,
  Boltzmanngasse 3, A-1090, Wien, Austria} \affiliation{Excellence
  Cluster Universe Technische Universit\"at M\"unchen, D-85748
  Garching, Germany}

\date{\today}

\begin{abstract}
This paper is addressed to the proof of the important role of
measuring apparatus, i.e. the measuring process, in the formation of
{\it necessary} and {\it sufficient} conditions for the explanation of
a time modulation of K--shell electron capture (EC) decay rates of
hydrogen--like (H--like) heavy ions (or the GSI oscillations) as the
interference of neutrino mass--eigenstates of the electron neutrino
constituents. For our analysis we use a toy--model, which has been
recently proposed by Peshkin arXiv:\,1403.4292 [nucl-th] for a
verification of the mechanism of the GSI oscillations as the
interference of neutrino mass--eigenstates by Ivanov and Kienle
Phys. Rev. Lett. {\bf 103}, 062502 (2009).
\end{abstract}

\maketitle

The GSI oscillations that is an unexpected time modulation of the
K--shell electron capture (EC) decay rates of the hydrogen--like
(H--like) ions have been observed at GSI Darmstadt in
\cite{GSI1}--\cite{GSI4} and confirmed recently in
\cite{Kienle2013}. In the GSI experiments on the EC decays $p \to d +
\nu_e$, where $p$ and $d$ are the parent H--like and daughter ions in
their ground states and $\nu_e$ is the electron neutrino, the rates of
the number of daughter ions $N_d(t)$ and the number of parent ions
$N_p(t)$ at time $t$ after the injection into the experimental storage
ring (ESR) are related by the equation
\begin{eqnarray}\label{label1}
\frac{dN_d(t)}{dt} = \lambda_{\rm EC}\,(1 + a \cos(\omega t + \phi))\,
N_p(t),
\end{eqnarray}
where $\lambda_{\rm EC}$ is the EC decay constant. The term $a
\cos(\omega t + \phi)$ defines an unexpected time modulation or the
GSI oscillations with an amplitude $a$, a period $T = 2\pi/\omega$ and
a phase--shift $\phi$. In turn, as has been reported in
\cite{GSI3,GSI4} and confirmed in \cite{Kienle2013} the rates of the
number of daughter ions $d'$ of the $\beta^+$--decays $p \to d' + e^+
+ \nu_e$ of the H--like heavy ions $p$, measured at the same
conditions as the EC decays $p \to d + \nu_e$, do not show a time
modulation.

As has been proposed in \cite{Ivanov2009a}--\cite{Ivanov2014}, the
time modulation of the EC decay rates and its unobservability in the
$\beta^+$--decay rates of the H--like heavy ions can be explained by
the interference of neutrino mass--eigenstates, which are constituents
of the electron neutrino defining the wave function of the electron
neutrino in the form of the superposition $|\nu_e\rangle =
\sum^{N_{\nu}}_{j = 1}U^*_{ej}|\nu_j\rangle$, where $U^*_{ej}$ are the
matrix elements of the $N_{\nu} \times N_{\nu}$ mixing matrix and
$N_{\nu} \ge 3$ is the number of neutrino mass--eigenstates
$|\nu_j\rangle$ with masses $m_j$ \cite{PDG12}. In such an approach
the EC decays $p \to d + \nu_e$ are defined by the decay channels $p
\to d + \nu_j$ ($j = 1,2,\ldots, N_{\nu}$), which may interfere at
certain conditions, caused by interactions of ions with measuring
apparatus in the ESR, and show the time modulation with frequencies
proportional to $(\Delta m^2_{ij})_{\rm GSI}/2M_p$, where $(\Delta
m^2_{ij})_{\rm GSI} = \tilde{m}^2_i - \tilde{m}^2_j$ are the
differences of squared dynamical masses $\tilde{m}_i$ and
$\tilde{m}_j$ of neutrino mass--eigenstates $|\nu_i\rangle$ and
$|\nu_j\rangle$, respectively, \cite{Ivanov2009a,Ivanov2009b} and
$M_p$ is the parent ion mass.  As has been shown in
\cite{Ivanov2009b}, neutrino mass--eigenstates in the weak decays of
highly charged heavy ions can acquire mass--corrections, caused by
polarisation of virtual $\nu_j \to \sum_{\ell}\ell^-W^+ \to \nu_j$
pairs in the strong Coulomb fields of daughter ions, where $\ell^-$
and $W^+$ are leptons and the $W^+$--boson of electroweak
interactions, respectively. These mass corrections increase the
differences $(\Delta m^2_{ij})_{\rm GSI}$, extracted from periods of
the GSI oscillations, in comparison to the differences $\Delta
m^2_{ij} = m^2_i - m^2_j$ of squared {\it bare} masses $m_j$ of
neutrino mass--eigenstates \cite{PDG12}. In contrast to the EC decay
rates the interference of the decay channels $p \to d' + e^+ + \nu_i$
and $p \to d' + e^+ + \nu_j$ ($i \neq j \in 1,2,\ldots, N_{\nu}$) of
the $\beta^+$--decays $p \to d' + e^+ + \nu_e$ occurs with much higher
frequencies $(\Delta m^2_{ij})_{\rm GSI}/2 Q_{\beta^+}$
\cite{Ivanov2008}, where $Q_{\beta^+}$ are the Q--values of the
$\beta^+$ decays, which are of order of a few MeV. As a result such a
time modulation is not observable in the GSI experiments
\cite{GSI3,GSI4,Kienle2013}.

Following \cite{Ivanov2009a}--\cite{Ivanov2014} one may state that i)
the {\it necessary} and {\it sufficient} condition for the
interference of neutrino mass--eigenstates is violation of energy and
3--momentum in the EC decays, caused by interactions of ions in the
ESR with measuring apparatus or by the measuring process
\cite{Wigner1963}, and ii) the orthogonality $\langle
d\nu_i|d\nu_j\rangle \sim \delta_{ij}$ of the final states of the
decay channels $p \to d + \nu_i$ and $p \to d + \nu_j$ ($i\neq j \in
1,2,\ldots, N_{\nu}$) of the EC decays $p \to d + \nu_e$ does not
prevent from the interference of neutrino mass--eigenstates.

Recently a toy--model for a verification of our explanation of the GSI
oscillations by means of the interference of neutrino
mass--eigenstates has been proposed by Murray Peshkin
\cite{Peshkin2014}. Within such a toy--model Peshkin confirms our
assertion that the orthogonality of the final states of the decay
channels $p \to d + \nu_i$ and $p \to d + \nu_j$ with $i \neq j \in
1,2,\ldots,N_{\nu}$ of the EC decay $p \to d + \nu_e$ makes no
influence on the suppression of the interference between neutrino
mass--eigenstates. However, his analysis of the suppression of the
interference of neutrino mass--eigenstates due to the absence of
off--diagonal matrix elements in the definition of the survival
probability of the parent ions suffers from the problem of the lack of
measuring apparatus and interactions of parent and daughter ions with
measuring apparatus, i.e. the measuring process \cite{Wigner1963}.

Let us repeat shortly Peshkin's analysis of our approach to the GSI
oscillations. According to Peshkin \cite{Peshkin2014}, the Hilbert
space of the vector states of the decaying system consists of the
parts in which the parent $p$ ion and decay products $d\nu_1$ and
$d\nu_2$ are present, where $d$ is a daughter ion and $|\nu_j\rangle$
are neutrino mass--eigenstates with masses $m_j$ for $j = 1,2$. Then,
let $|\psi(t)\rangle$ be the vector state of the entire system and let
$P_p$ be the projection operator on the parent ion state
$|P_p\psi(t)\rangle$. The translational invariant Hamilton operator of
the system is given by ${\rm H} = {\rm H}_0 + V$, where ${\rm H}_0$ is
the Hamilton operator of the free objects $p$, $d$ and $\nu_j$ for $j
= 1,2$, whereas $V$ describes the interaction between them. The
survival probability $S(t)$ of the parent ion at time $t$ after the
injection into the ESR is defined by \cite{Peshkin2014}
\begin{eqnarray}\label{label2}
S(t) = \langle P_p\psi(t)|P_p\psi(t)\rangle = \langle\psi(0)|e^{+ i
  {\rm H}t}P_pe^{-i {\rm H}t}|\psi(0)\rangle.
\end{eqnarray}
The rate $R(t)$ of the survival probability is given by
\begin{eqnarray}\label{label3} R(t) = - \frac{dS(t)}{dt}
 \propto 1 + a\cos(\omega t + \phi),
\end{eqnarray}
where the time modulated term $a\cos(\omega t + \phi)$ can appear only
if the interference of the decay channels $p \to d + \nu_1$ and $p \to
d + \nu_2$ or neutrino mass--eigenstates $|\nu_1\rangle$ and
$|\nu_2\rangle$ is allowed.

For the analysis of the conditions, at which the interference between
the decay channels $p \to d + \nu_1$ and $p \to d + \nu_2$ can occur,
Peshkin transcribes the survival probability into the form
\begin{eqnarray}\label{label4}
S(t) = \langle\psi(0)|e^{+ i {\rm H}t}P_pe^{-i {\rm H}t}|\psi(0)\rangle = \int
d^3K'd^3K\,\langle \psi(0)|\vec{K}\rangle\langle \vec{K}\,|e^{+ i
  {\rm H}t}P_pe^{-i {\rm H}t}|\vec{K}\,'\rangle\langle \vec{K}\,'|\psi(0)\rangle,
\end{eqnarray}
where $\vec{K}$ and $\vec{K}\,'$ are total 3--momenta and other
dynamical variables accompanying $\vec{K}$ and $\vec{K}\,'$ are
assumed \cite{Peshkin2014}. The interference of the decay channels $p
\to d + \nu_1$ and $p \to d + \nu_2$ or the time modulation can appear
in the survival probability $S(t)$ only due to off--diagonal matrix
elements $\langle \vec{K}\,|e^{+ i {\rm H}t}P_pe^{-i {\rm
    H}t}|\vec{K}\,'\rangle \neq 0$ \cite{Peshkin2014}. However, as has
been pointed out by Peshkin, the matrix elements $\langle
\vec{K}\,|e^{+ i {\rm H}t}P_pe^{-i {\rm H}t}|\vec{K}\,'\rangle$ should
be diagonal
\begin{eqnarray}\label{label5}
\langle \vec{K}\,|e^{+ i {\rm H}t}P_pe^{-i {\rm H}t}|\vec{K}\,'\rangle
\sim \delta^{(3)}(\vec{K} - \vec{K}\,')\,\langle \vec{K}\,|e^{+ i {\rm
    H}t}P_pe^{-i {\rm H}t}|\vec{K}\,\rangle,
\end{eqnarray}
since the Hamilton operator ${\rm H}$ is translational
invariant. Peshkin also assumes that off--diagonal matrix elements can
appear due to external fields such as magnetic fields, stabilising a
motion of ions in the ESR, and electron cooling in the ESR. However,
such a possibility has been rejected by Peshkin, since in this case
parameters of a time modulation should depend on parameters,
characterising these fields, and a frequency of a time modulation may
or may not depend on $\Delta m^2_{21} = m^2_2 - m^2_1$, where $m_2$
and $m_1$ are the masses of neutrino mass--eigenstates $|\nu_2\rangle$
and $|\nu_1\rangle$, respectively.

Below we show that Peshkin's analysis of off--diagonal matrix elements
of the survival probability $S(t)$, based on the use of the
translational invariant Hamilton operator ${\rm H}$ for the
description of evolution of the decaying system and its decay products
in the ESR, is not complete and interactions of ions with measuring
apparatus, i.e. the measuring process, should be taken into account
\cite{Wigner1963}. Indeed, apart from the magnetic fields, stabilising
a motion of ions in the ESR after the injection of parent ions into
the ESR and cooling, that reduces a relative transverse velocity
spread to $\Delta v/v \approx 5\times 10^{-7}$ \cite{Kienle2013}, ions
are being monitored by both the resonant pickup (the 245 MHz
resonator) and the capacitive pickup (the Schottky noise detector)
during an optimised measuring period $54\,{\rm s}$ with sampling times
$\delta t = 64\,{\rm ms}$ and $\delta t = 32\,{\rm ms}$, respectively
\cite{Kienle2013}. This implies that evolution of ions in the ESR
cannot be analysed correctly if interactions of ions with measuring
apparatus is switched off, i.e. using only the Hamilton operator ${\rm
  H}$.

\begin{figure}
\centering
\includegraphics[width=0.4\linewidth]{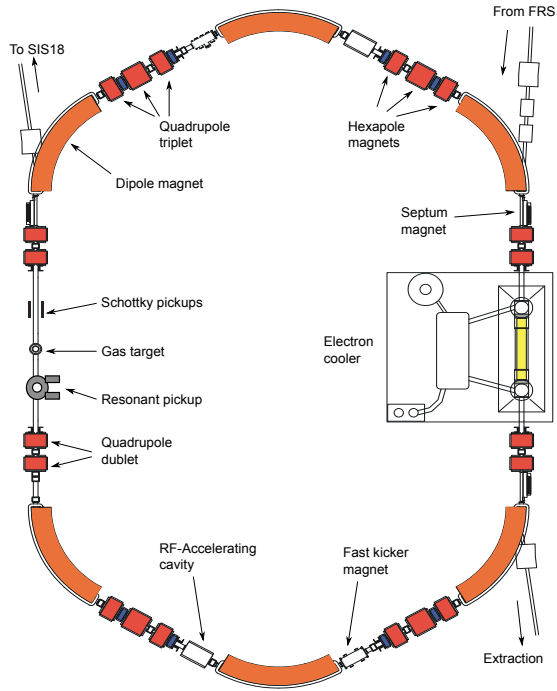}
\caption{The ESR at GSI, equipped by the resonant and capacitive
  pickups detecting parent and daughter ions \cite{Shahab2013}.}
 \label{fig:esr}
\end{figure}

The account for interactions between measuring apparatus and a quantum
system leads to a change of its wave function. According to Wigner
\cite{Wigner1963}, von Neumann \cite{Neumann1932} and London and Bauer
\cite{London1939} (see also \cite{Wheeler1983} and
\cite{Caves1986,Caves1987,Caves1987a}), wave functions of quantum
systems, coupled to measuring apparatus, evolve in time with i)
unitary Schr\"odinger evolution between measurements and ii) sudden
and non--unitary evolution at time of each measurement. Moreover for
sequence of measurements, distributed in time, measurements can
overlap and interleave in a complicated way \cite{Caves1987}. As a
result, there can be in general no time, when quantum systems are
undisturbed by measurements, and, correspondingly, no time intervals
of unitary evolution \cite{Caves1987}.

How such a dynamics of evolution of wave functions of highly charged
heavy ions, coupled to measuring apparatus, is realised in GSI
experiments?  In the ESR ions, moving after cooling with a velocity $v
= 0.71$, make a total revolution over the circumference of the ESR
with the length $\Pi = 108.36\,{\rm m}$ for $T_{\rm rev} = 510\,{\rm
  ns}$. During the measurement time $t \le 54\,{\rm s}$ ions make
enormous number of total revolutions, going during every revolution
through the resonant and capacitive pickups, located at the ESR as it
is shown in Fig.\,{\ref{fig:esr}}.

Passing through the resonant and capacitive pickups ions lose and gain
energy \cite{Shahab2013}. Such a process has a random and cumulative
character, which leads to energy violation and to disturbance of wave
functions of ions.  For example, in the resonant pickup the
interaction of ions with only the fundamental mode leads to an energy
loss of order $\Delta {\cal E} \sim 15\,{\rm \mu eV}$
\cite{Shahab2013}. Such an energy loss corresponds to an observation
time $\Delta t \sim 2\pi/\Delta {\cal E} \sim 0.28\,{\rm ns}$
\cite{Neumann1932,Bohm1961}, during which ions are observed
continuously \cite{Shahab2013}. Thus, during the observation time
$\Delta t \sim 0.27\,{\rm ns}$, caused by the interaction with only
the fundamental mode of the resonant pickup, wave functions of ions
evolve according to non--unitary evolution. In the capacitive pickup
ions pass through a pair of parallel metal plates and induce images of
charges on the surface of each plate \cite{Frock2012}. The signals
coming from the two plates are amplified and used in dependence on
specific purposes \cite{Shahab2013}. Interactions of ions with their
images and plates lead to energy fluctuations and to disturbance of
wave functions ions and, correspondingly, to non--unitary
evolution. It is obvious that wave functions of ions cannot be
restored to their undisturbed form immediately after the exit from the
region of the resonant and capacitive pickups and there is a certain
retardation of such a restoration, which, of course, becomes slower
and slower with the number of revolutions of ions over the
circumference of the ESR.  As a result, the resonant and capacitive
pickups divide the ESR into the regions of non--unitary and unitary
evolution of wave functions of ions. Such a decomposition of the ESR
is not stationary. Time intervals of non--unitary and unitary
evolution increase and decrease, respectively. Of course, a detection
of ions by both the resonant and capacitive pickups does not destroy
the quantum states of H--like heavy ions, since energy fluctuations
are much smaller then binding energies, but causes a smearing of
energies and momenta of daughter ions $d$ leading to
indistinguishability of daughter ions in the decay channels $p \to d +
\nu_1$ and $p \to d + \nu_2$ and, correspondingly,
indistinguishability of the decay channels $p \to d + \nu_1$ and $p
\to d + \nu_2$ of the EC decay $p \to d + \nu_e$ leading to their
interference \cite{Ivanov2009a}--\cite{Ivanov2014}. As has been shown
in \cite{Ivanov2009a}--\cite{Ivanov2014}, indistinguishability of
daughter ions and the decay channels $p \to d + \nu_1$ and $p \to d +
\nu_2$ can be qualitatively described in terms of Heisenberg's
uncertainty relations.

Hence, for the correct analysis of the origin of the interference of
neutrino mass--eigenstates the Hamilton operator of Peshkin's
toy--model should be taken in the form ${\rm H}_{\rm tot} = {\rm H} +
{\rm U}(t)$ \cite{Caves1987,Caves1986,Caves1987a,Bohm1961}, where
${\rm U}(t)$ is a time--dependent and translational non--invariant
potential of ions coupled to measuring apparatus, i.e. the resonant
and capacitive pickups. The potential ${\rm U}(t)$ does not vanish
only during time intervals of observation of parent and daughter ions,
coupled to the resonant and capacitive pickups \cite{Bohm1961}. A time
evolution of the Hamilton operator ${\rm H}$ and the 3--momentum
operator $\vec{K}$ of parent ions and decay products is defined by the
Heisenberg equations of motion \cite{Caves1987a}
\begin{eqnarray}\label{label6}
\frac{d{\rm H}}{dt} = i\,[{\rm U}(t),{\rm H}]\quad,\quad
\frac{d\vec{K}}{dt} = i\,[{\rm U}(t),\vec{K}\,].
\end{eqnarray}
Since $[{\rm U}(t), {\rm H}] \neq 0$ and $[{\rm U}(t), \vec{K}\,] \neq
0$, the Heisenberg equations of motion testify that energy and
momentum in the EC decays are violated \cite{Davydov1965}. This
results in non--vanishing off--diagonal matrix elements in
Eq.(\ref{label4}) and provides the basis for the time modulation of
the survival probability $S(t)$ and its rate $R(t)$.  In other words
violation of energy and momentum in the EC decays, caused by
interactions of parent and daughter ions with measuring apparatus,
defines {\it necessary} and {\it sufficient} conditions for the
appearance of the time modulation of the EC decay rate $p \to d +
\nu_e$ as the interference of the decay channels $p \to d + \nu_1$ and
$p \to d + \nu_2$ or neutrino mass--eigenstates $|\nu_1\rangle$ and
$|\nu_2\rangle$ \cite{Ivanov2014}. A proportionality of the modulation
frequency to $\Delta m^2_{21}/2 M_p = (m^2_2 - m^2_1)/2 M_p$ is
justified by non--conservation and smearing of 3--momenta $\vec{q}_1$
and $\vec{q}_2$ of the daughter ions in the decay channels $p \to d +
\nu_1$ and $p \to d + \nu_2$ around a 3--momentum $\vec{q}\pm \delta
\vec{q}$, where $\delta \vec{q}$ is uncertainty of 3--momentum of
daughter ions caused by the measuring process
\cite{Ivanov2009a,Ivanov2010a,Ivanov2010b,Ivanov2014}.

One of us (A.N.I.) is grateful to Murray Peshkin for calling attention
to the paper \cite{Peshkin2014} and Shahab Sanjari for giving Fig.\,1
and numerous fruitful discussions on the properties of the resonant
and capacitive pickups and on GSI experiments on a time modulation of
weak decay rates of highly changed heavy ions. A.N.I. thanks also
Shahab Sanjari for reading the manuscript and useful comments. This
work was supported by the Austrian ``Fonds zur F\"orderung der
Wissenschaftlichen Forschung'' (FWF) under the contract I862-N20.

\end{document}